\documentclass[twocolumn,showpacs,preprintnumbers,amsmath,amssymb]{revtex4}

\usepackage{graphicx}
\usepackage{dcolumn}
\usepackage{bm}

\begin{document}

\preprint{APS/123-QED}

\title{  Enhanced transmission of transverse electric waves through subwavelength slits in a thin metallic film}

\author{Yu Qian Ye$^{1,2}$ }
\author{Yi Jin$^1$}
 \email{jinyi@coer.zju.edu.cn}
\affiliation{%
$1.$Centre for Optical and Electromagnetic Research, and Joint
Research
Centre of Photonics of \\
the Royal Institute of Technology (Sweden) and Zhejiang University,
Zijingang Campus, Zhejiang University, China \\
$2.$ Department of Physics, Zhejiang University, Hangzhou 310027,
China
}%

\date{\today}

\begin{abstract}
By coating a cover layer with metallization of cut wire array, the
transmission of transverse electric waves (TE; the electric field is
parallel to the slits) through subwavelength slits in a thin
metallic film is significantly enhanced. An 800-fold enhanced
transmission is obtained compared to the case without the cut wires.
It is demonstrated that a TE incident wave is highly confined by the
cut wires due to the excitation of the electric dipole-like
resonance, and then effectively squeezed into and through the
subwavelength slits.
\end{abstract}

\maketitle

Since the first experimental report of ``extraordinary optical
transmission'' (EOT) through an metallic grating with periodic
subwavelength holes \cite{Porto}, there has been considerable
interest in the ``enhanced" transmission phenomena of subwavelength
apertures in metals, such as one-dimensional (1D) periodic arrays of
slits \cite{Porto,Cao}, two-dimensional (2D) periodic arrays of
cylindrical \cite{Ebbesen,Mart} and rectangular (square) holes
\cite{Molen,Klein}, and a single aperture surrounded by periodic
corrugations \cite{Takakura,Moreno,Vidal}. It is worth noting that
there are two different transmission mechanisms of subwavelength
apertures. Since the subwavelength holes do not support the
propagating modes, the transmission of incident waves can only rely
on the evanescent tunneling process that can be enhanced by either
diffraction or surface wave excitation. As to the subwavelength
slits, there is always a propagating waveguide mode with no cutoff
frequency \cite{Kong} for transverse magnetic (TM) waves (i.e., the
magnetic field parallel to the slits). Utilizing this guided mode
can result to strong transmission of TM waves through the slits
\cite{Went}.

Lots of works on the EOT phenomena of subwavelength slits have been
reported in microwave and optical frequencies
\cite{Bravo,Lockyear,Yang}.  However, most of them have been focused
on TM waves. This paper is dedicated to enhance the transmission of
TE waves (i.e., the electric field is parallel to the slits) through
subwavelength slits in a thin metallic film. Different from TM
waves, the transmission of a TE wave through a subwavelength slit
has a cut-off frequency \cite{Kong}, below which no propagating mode
exists. By coating a cover layer with a metallization of cut wire
array, the transmission of a TE wave through an array of
subwavelength slits can be drastically enhanced at a resonant point
below the cut-off frequency.

\begin{figure}
\resizebox{0.5\textwidth}{!}{
\includegraphics{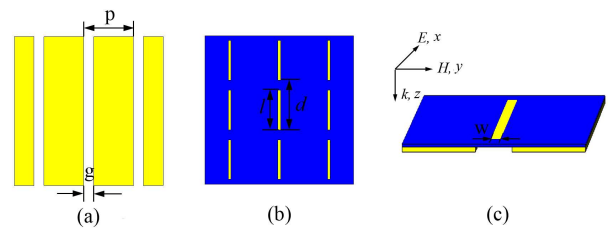}}
\caption{\label{Fig1} (a) The top view of 1D periodic slit array
perforated in a metallic film. (b) the top view of a cover layer
with a metallization of cut wire array. (c) Unit cell for
transmission calculation when the cover layer is put  perforated
film in (a) to enhance the transmission. Axes indicate the
propagation direction and polarization of the incident plane wave.}
\end{figure}

\begin{figure}
\resizebox{0.5\textwidth}{!}{
\includegraphics{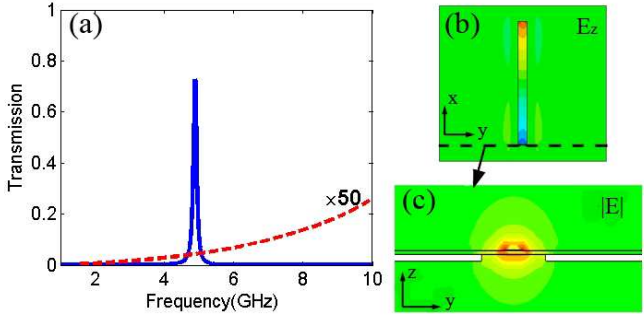}}
\caption{\label{Fig2} (color online) (a) Transmission of a slit
array with (blue solid curve) and without (red dashed) the cover
layer shown in Fig. 1(b). (b) Distribution of the z-component
electric field on the cut wire at the resonant frequency. (c)
Distribution of the electric field magnitude near the aperture of
the slit in the y-z plane at the side edge of the cut wire [see the
construction line in Fig. 2(b)]. }
\end{figure}

A 1D periodic array of subwavelength slits perforated in an aluminum
film is shown in Fig. $\ref{Fig1}$(a). The thickness of the metallic
film is $t=0.5$ ${\rm mm}$, the width of the slits is $g=5.5$ ${\rm
mm}$, and the period of the slit array is $p=30$ ${\rm mm}$ (these
parameters are chosen to be practical in experiments). By using a
Finite-Integration Time Domain algorithm \cite{CST}, the
transmission spectrum of the slit array is calculated and shown by
the dashed (red) curve in Fig. $\ref{Fig2}$(a) (where the
transmission values are amplified by 50 times to improve the
visibility). In calculation, a TE plane wave is normally incident on
the structure. Since the considered wavelength ($\lambda>30$ ${\rm
mm}$) is much larger than the slit width ($g<\lambda/5$) no
propagating mode can be excited by the TE plane wave  \cite{Kong}.
So the transmission can only rely on the evanescent tunneling
process, and is pretty low ($<0.1\%$) even for such thin metallic
film at microwave frequencies.

\begin{figure}
\resizebox{0.45\textwidth}{!}{
\includegraphics{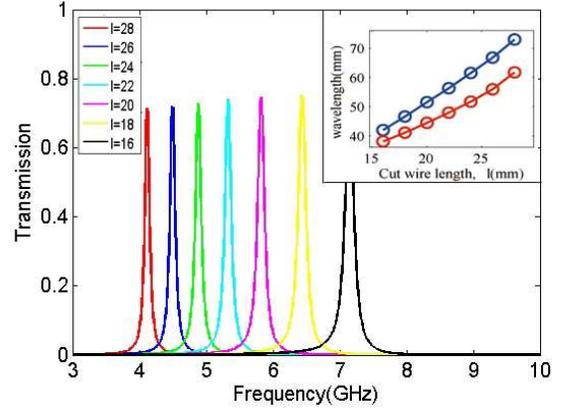}}
\caption{\label{Fig3} (color online). Transmission of the composite
structure in Fig. $\ref{Fig1}$(c) for different cut-wire length $l$.
The inset shows the resonant wavelength as a function of the
cut-wire length $l$ for the composite structure (blue curve) and
only the cover layer in Fig. $\ref{Fig1}$(b) (red curve). }
\end{figure}

To enhance the transmission of the subwavelength slits, a cover
layer which is formed of cut wire array metallized on a dielectric
substrate is used, as shown in Fig. $\ref{Fig1}$(b). The
metallization was in copper with a thickness of $0.02$ ${\rm mm}$.
The cut wires are of length $l=24$ ${\rm mm}$ and width $w_1=1.6$
${\rm mm}$. The period of the cut-wire array in the x and y
directions are $p=30$ ${\rm mm}$ and $d=30$ ${\rm mm}$,
respectively. The dielectric substrate is of permittivity
$\varepsilon=3.0$ and thickness $t_1=0.3$ ${\rm mm}$. The unit cell
of the composite structure for transmission calculation is shown in
Fig. $\ref{Fig1}$(c). As the simulation result represents (see the
solid curve in Fig. $\ref{Fig2}$(a)), a resonant transmission peak
appears at frequency $\omega=4.88$ ${\rm GHz}$ ($\lambda=61.5$ ${\rm
mm}$) with transmissivity of up to $72\%$, which is about $800$
times larger than that in the case without the cover layer. At
resonance, the distribution of the z-component electric field on the
cut wire is shown in Fig. $\ref{Fig2}$(b), which represents that
charges of opposite signs accumulate at the two ends of the broken
wire, indicating the excitation of a strong electric dipole-like
resonance \cite{Zhou} on the cut wire. The distribution of the
corresponding electric field magnitude on the y-z plane at the side
edge of the cut wire is shown in Fig. $\ref{Fig2}$(c), from which
one sees that the resonant electric field is highly localized in a
subwavelength range around the cut wire. Based on the strong
localized electric field near the input apertures of the slits, the
incident wave is effectively coupled into evanescent waves, and then
squeezed through the subwavelength slits. Consequently, a striking
resonant transmission is obtained. To further demonstrate that the
drastic transmission improvement is caused by the electric
dipole-like resonance of the cut wires, Fig. $\ref{Fig3}$ shows the
transmission for different cut-wire length $l$, where $l$ varies
from $16$ ${\rm mm}$ to $28$ ${\rm mm}$. The resonant frequency of
the transmission peak obviously shifts with $l$, while the peak
amplitude changes little. The corresponding resonant wavelength of
the transmission peak is shown by the blue curve in the inset of
Fig. $\ref{Fig3}$ as a function of $l$. For comparison, the resonant
wavelength of the fundamental mode (i.e., the electric dipole-like
resonant mode) of the cut wires for different $l$ is also shown in
the inset of Fig. $\ref{Fig3}$ (red curve). The correlation of the
two curves indicates that the resonant transmission frequency is
dominantly determined by the resonance of the cut wires. The
relatively small frequency difference between the two curves is
caused by the interaction of the cut wires with the patterned
metallic film.

\begin{figure}
\resizebox{0.5\textwidth}{!}{
\includegraphics{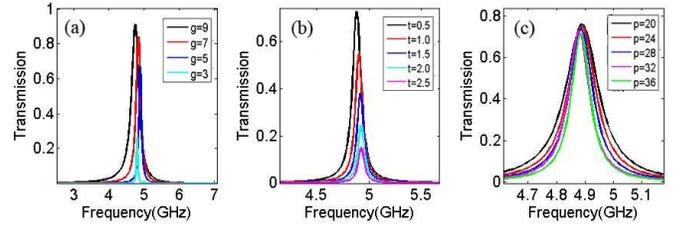}}
\caption{\label{Fig4} (Color online). Transmission spectra of the
composite structure in Fig. $\ref{Fig1}$(c). The parameters are the
same as those in Fig. $\ref{Fig2}$(a), expect that (a) $g$ varies
from $9$ to $3$ ${\rm mm}$; (b) $t$ varies from $0.5$ to $2.5$ ${\rm
mm}$; (c) $p$ varies from $20$ to $36$ ${\rm mm}$.}
\end{figure}

Next, the influence of the geometric parameters of the slit array on
the resonant transmission of the composite structure is
investigated. Figs. 4(a) and 4(b) show the corresponding
transmission when the thickness $t$ of the metallic film and the
width $g$ of the slits are changed, respectively. The other
parameters are the same as those used in Fig. $\ref{Fig2}$. The
amplitude of the resonant transmission peak rapidly drops with the
decrease of $g$ and increase of $t$. It is not unexpected that
reducing the size of the apertures and thickening the metallic film
will reduce the evanescent wave tunneling and diminish the
transmission, like the case of surface plasmon enhanced EOT
\cite{Degiron,van} through the subwavelength hole array.
Nevertheless, the current enhanced transmission is not due to
surface plasmon, as the incident wave is a TE wave and the structure
works at microwave frequencies. In Figs. 4(a) and 4(b), a striking
point is that the peak frequency does not obviously shift with the
variation of $t$ and $g$, because the transmission is based on the
evanescent tunneling process and the resonant frequency is
determined by the cut wires as discussed in the above. Notably, in
the previous works on the EOT phenomena of TE waves \cite{Lu,Crouse}
through the slit which is need to be wide enough to support
propagating modes, the frequency of resonant transmission is
sensitive to the geometric parameters of slits, since the EOT is
caused by the excitation of cavity modes in the slits. Fig.
$\ref{Fig4}$(c) shows the transmission of the composite structure
when the period $p$ is varied. The transmission peak is not
perceptibly changed with the variation of $p$, which indicates that
Bloch surface waves \cite{Ruan} are not contributed to the
drastically enhanced transmission

\begin{figure}
\resizebox{0.5\textwidth}{!}{
\includegraphics{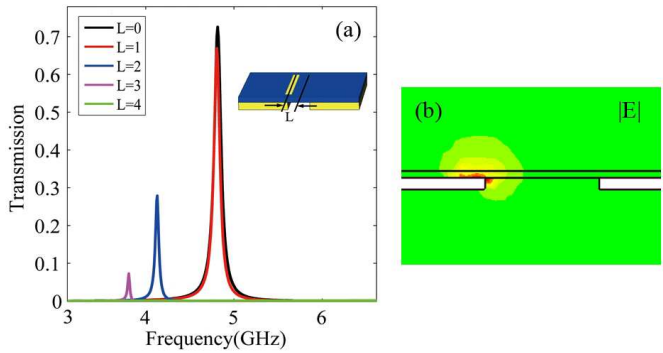}}
\caption{\label{Fig5} (color online). (a) Transmission of the
structure shown in the inset. The parameters are the same as those
used in Fig. $\ref{Fig2}$(a) except $L$. (b) Distribution of the
electric field magnitude in the y-z plane at the side edge of the
cut wire for $L=3$ ${\rm mm}$.}
\end{figure}

It is also interesting to explore the transmission behavior when the
cover layer is laterally displaced relative to the patterned
metallic film. The lateral shift is denoted by $L$ in the inset of
Fig. $\ref{Fig5}$(a). The other parameters are the same as those
used in Fig. $\ref{Fig2}$. The transmission for different $L$ is
shown in Fig. $\ref{Fig5}$(a). When $L$ is below $1.0$ ${\rm mm}$,
the transmission does not perceptibly change [see the red curve in
Fig. $\ref{Fig5}$(a)]. That is to say that the resonant transmission
is quite stable for small lateral displacement which could be easily
induced by experimental error. As $L$ is further increased, the
amplitude of the resonant transmission peak rapidly drops. When $L$
reaches $4.0$ ${\rm mm}$, no resonant transmission peak is observed
as shown by the green line (overlap with the x-axis) in Fig.
$\ref{Fig5}$(a). The distribution of the electric field magnitude in
Fig. $\ref{Fig5}$(b) can well explain the above behavior. The
electric dipole-like resonance is highly localized around the cut
wire. When $L$ is small, the localized resonant field is blocked
little by the metallic film, and the transmission is insignificantly
influenced. When $L$ becomes large, a part of the localized resonant
field is blocked by the metallic film, and the resonant transmission
is suppressed since the incident wave can not effectively coupled
into the slits. As shown by the above simulation results, this
resonant transmission enhancement can be manipulated just by
changing the laterally displacement $L$.

In conclusion, we have numerically demonstrated that at microwave
frequencies the transmission of TE waves through subwavelength slits
in a thin metallic film can be greatly enhanced by assistant cut
wires. The electric dipole-like resonance which makes the incident
TE wave tightly confined around the cut wires plays a key role in
the enhanced transmission mechanism. Although our numerical
calculation were carried out at microwave frequencies, the proposed
method of transmission enhancement can be easily extended to
terahertz frequencies or optical frequencies due to the simplicity
of the present structure.

This work is partly supported by the National Basic Research Program
(No. 2004CB719801), the National Natural Science Foundations (NNSF)
of China under Project No. 60688401 and 60677047.


\end{document}